\documentclass[%
 reprint,
 superscriptaddress,showkeys,
 amsmath,amssymb,
 aps,
]{revtex4-1}

\usepackage{graphicx}
\usepackage{dcolumn}
\usepackage{bm}
\usepackage{xcolor}
\usepackage{url}
\usepackage{makecell}
\usepackage{subfigure}
\usepackage{multirow}
\usepackage{booktabs}

\begin{document}


\title{Exclusive vector meson productions with the analytical solution of Balitsky-Kovchegov Equation}

\author{Xiaopeng Wang}
\email{wangxiaopeng@impcas.ac.cn}
\affiliation{Institute of Modern Physics, Chinese Academy of Sciences, Lanzhou 730000, China}
\affiliation{Lanzhou University, Lanzhou 730000, China}
\affiliation{University of Chinese Academy of Sciences, Beijing 100049, China}

\author{Wei Kou}
\affiliation{Institute of Modern Physics, Chinese Academy of Sciences, Lanzhou 730000, China}
\affiliation{University of Chinese Academy of Sciences, Beijing 100049, China}

\author{Gang Xie}
\affiliation{Guangdong Provincial Key Laboratory of Nuclear Science, Institute of Quantum Matter, South China Normal University, Guangzhou 510006, China}
\affiliation{Institute of Modern Physics, Chinese Academy of Sciences, Lanzhou 730000, China}

\author{Ya-Ping Xie}
\email{xieyaping@impcas.ac.cn }
\affiliation{Institute of Modern Physics, Chinese Academy of Sciences, Lanzhou 730000, China}

\author{Xurong Chen}
\email{xchen@impcas.ac.cn}
\affiliation{Institute of Modern Physics, Chinese Academy of Sciences, Lanzhou 730000, China}
\affiliation{Guangdong Provincial Key Laboratory of Nuclear Science, Institute of Quantum Matter, South China Normal University, Guangzhou 510006, China}
\affiliation{School of Nuclear Science and Technology, University of Chinese Academy of Sciences, Beijing 100049, China}

\date{\today}

\begin{abstract}
Exclusive vector meson production is an excellent probe for describing the structure of proton. In this paper, based on dipole model, the differential cross sections, total cross sections and the ratios of the longitudinal to transverse cross sections of $J/\psi$ and $\rho^0$ productions are calculated with the analytical solution of Balitsky-Kovchegov (BK) equation. In addition, we also consider the influences of two meson wave function models on the results. Our predictions, which are little sensitive to meson wave functions, agree with the experimental data. The analytical solution of BK equation is reliable for description of exclusive vector meson production in a certain range of $Q^2$.
\end{abstract}

\keywords{exclusive vector meson production, Balitsky-Kovchegov equation, analytic solution}
\pacs{14.40.-n,  13.60.Hb, 13.85.Qk}

\maketitle

\section{Introduction}
\label{sec:intro}

 Color Glass Condensate (CGC) \cite{McLerran:1993ni,Iancu:2002xk,Iancu:2003xm,Jalilian-Marian:2005ccm,Gelis:2010nm} is an effective theory to describe physics in the proton saturation regime. Inside proton, gluons cannot grow infinitely without breaking the unitary limit. Consequently, in small x regime, recombination and multiple scattering  of gluon will reach the balance, then the number of gluons will not increase. This is called gluon saturation, which can be described by CGC effective field theory. For studying proton structure in saturation regime at high energy limit, deep inelastic scattering (DIS), deeply virtual compton scattering (DVCS) and exclusive diffractive vector meson production are excellent probes \cite{Gribov:1983ivg,Mueller:1985wy,Mueller:1999wm}.
 
 For analyzing DIS and vector meson production, the color dipole model or CGC effective theory is a powerful tool \cite{Nikolaev:1990ja,Mueller:1994jq,Mueller:1993rr,Kopeliovich:1991pu,Nikolaev:1990ja,Kopeliovich:1993pw,Nemchik:1994fp,Nemchik:1996pp,Nemchik:1996cw}. In dipole model, virtual photon splits into a quark anti-quark pair (dipole) which interacts with proton by exchanging gluons. Then the dipole recombines into the meson or the photon. Golec-Biernat and Wusthoff (GBW) model \cite{Golec-Biernat:1998zce} and CGC model \cite{Iancu:2003ge} have a good description of the dipole scattering process. And two impact-parameter dependent models, b-CGC model \cite{Watt:2007nr,Kowalski:2006hc} which is the modification of CGC model and \textbf{IP}-Sat \cite{Kowalski:2003hm,PhysRevD.87.034002} model are successful in the application of DIS and exclusive vector meson production process \cite{Iancu:2003ge,Lappi:2010dd,Kowalski:2007rw,Kowalski:2008sa,Xie:2022sjm}. Saturation effect in b-CGC model and \textbf{IP}-Sat model is related to Balitsky-Kuraev-Fadin-Lipatov (BFKL) equation \cite{Fadin:1975cb,Kuraev:1976ge,Kuraev:1977fs,Balitsky:1978ic} and Dokshitzer-Gribov-Lipatov-Altarelli-Parisi (DGLAP) equation \cite{Gribov:1972ri,Dokshitzer:1977sg,Altarelli:1977zs} respectively \cite{Rezaeian:2013tka}. These models are also applied to Large Hadron Collider(LHC) experiment \cite{Motyka:2008ac,Levin:2010dw,Tribedy:2010ab,Tribedy:2010ab,Xie:2018rog}.      
 
 Besides, the evolution of dipole-target scattering amplitude can be dominated by 
  Balitsky-Kovchegov (BK) equation \cite{Balitsky:1997mk,Kovchegov:1999yj,Kovchegov:1999ua,Balitsky:2001re} with a nonlinear term for gluon saturation, which is regarded as the mean field approximation of the Jalilian-Marian-Iancu-McLerran-Weigert-Leonidov-Kovner (JIMWLK) equation \cite{Balitsky:1995ub,Jalilian-Marian:1997qno,Iancu:2000hn,Weigert:2000gi}. There are a lot of research on the BK equation.
 The influence of impact parameter b was numerically analyzed for BK equation in \cite{Golec-Biernat:2003naj}. The solution of the b-dependence BK  equation with the collinearly improved kernel has been studied \cite{Bendova:2019psy}. Discussion for the behavior
of the numerical solution of running coupling BK equation with Runge-Kutta method was presented in \cite{Matas:2016bwc}. Re.$\,$\cite{Lappi:2015fma} showed numerical solution to BK equation in the coordinate space at next-to-leading order.

In the momentum space, it has been shown \cite{Kovchegov:1999yj,Kovchegov:1999ua} that the nonlinear evolution equation with the Balitskii-Fadin-Kuraev-Lipatov (BFKL) kernel is obtained form BK equation by proper approximation. Then, by variable substitution, this nonlinear evolution equation \cite{Munier:2003vc,Munier:2003sj,Munier:2004xu} is simplified to Fisher-Kolmogorov-Petrovsky-Piscounov (FKPP) equation \cite{fisher1937the} which is a kind of reaction-diffusion nonlinear equation. The analytical solution of BK equation is obtained by solving the more concise FKPP equation. The analytical solution \cite{Bondarenko:2015fca,Marquet:2005ic,Xiang:2017fjr,Xiang:2019kre,Xiang:2020xxe} of BK equation has been studied by different methods. We also obtain the analytical solution of BK equation \cite{Wang:2020stj} in the momentum space.
In this work, the analytical solution that we have obtained will be used to explain the scattering process between protons and color dipole.

The arrangement of the article is as follows. In Sec.$\,$\ref{sec:DIS and DVMP}, the dipole representation of exclusive vector meson production is reviewed, and overlaps between photon and vector mesons ($J/\Psi$, $\rho^0$) wave function are shown. in Sec.$\,$\ref{sec:solution of BK equation}, Our solution of BK equation in momentum space is introduced. And the results of the fitting to the structure function $F_2$ of proton with our solution are provided. In Sec.$\,$\ref{sec:Numerical results}, the figures of our calculations for vector mesons ($J/\Psi$, $\rho^0$) production differential cross section with $\rm t$ and total cross section as functions of center of mass energy W and the photon virtuality $Q^2$ are presented. The ratios of the longitudinal to transverse cross section for $J/\Psi$ and $\rho^0$ are shown. Finally, in Sec.$\,$\ref{sec:summary}, discussion and summary are given.

\section{Dipole description of exclusive vector meson production }
\label{sec:DIS and DVMP}
\begin{figure*}
	\begin{minipage}[t]{0.5\linewidth}
		\centering
		\includegraphics[scale=0.5]{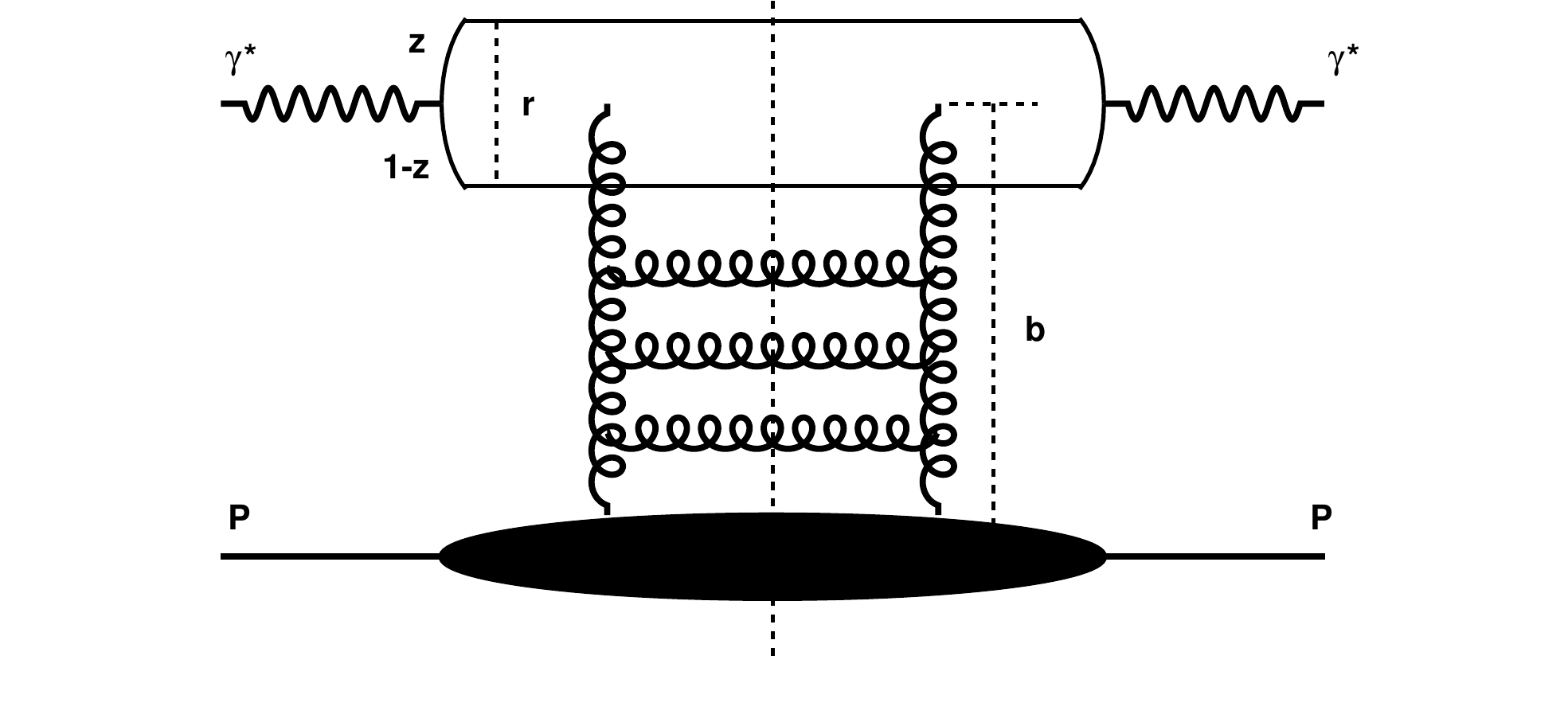}
	\end{minipage}%
	\begin{minipage}[t]{0.5\linewidth}
		\centering
		\includegraphics[scale=0.5]{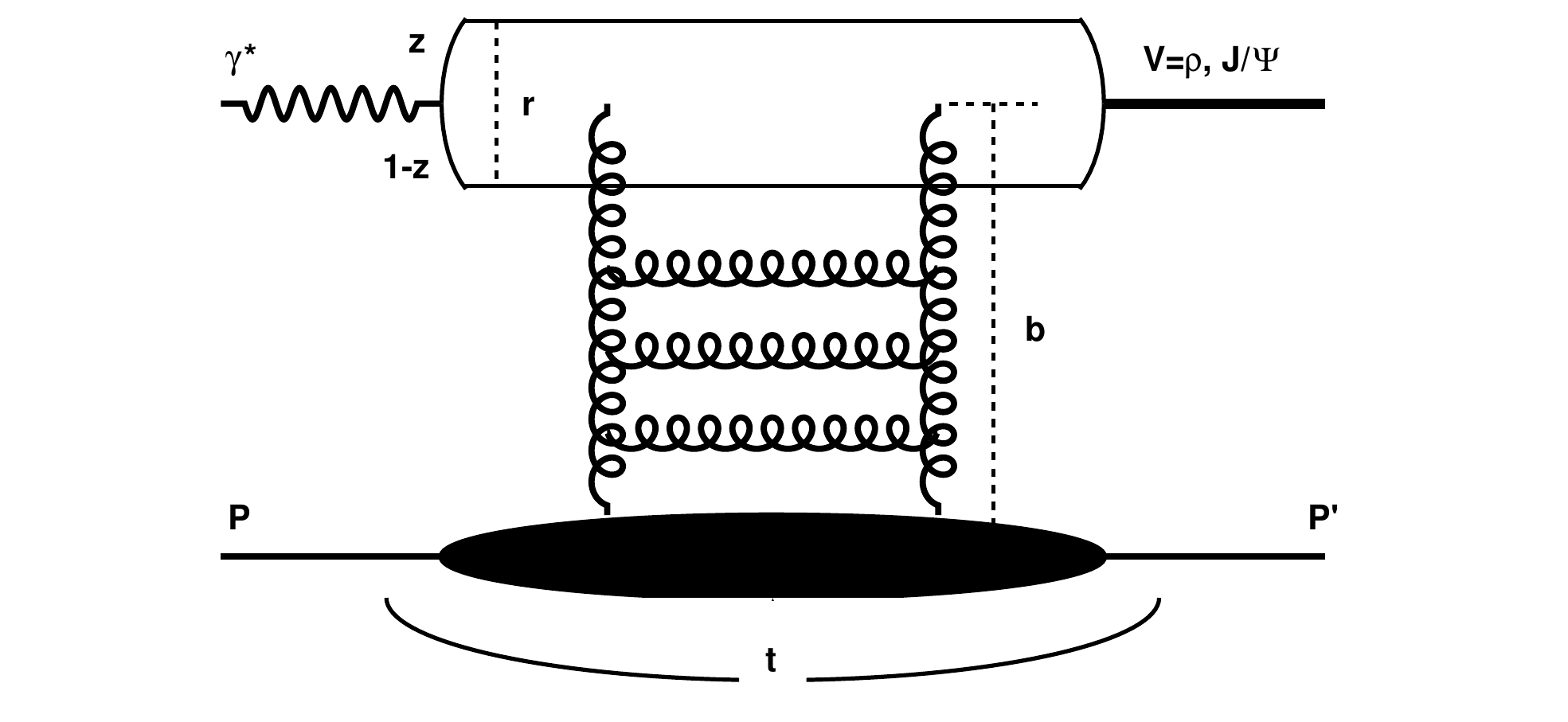}
	\end{minipage}
	\caption{(Left) Elastic scattering amplitude for DIS and (Right) the amplitude of exclusive vector meson production in dipole model.}
	\label{fig:DISandDVMP}
\end{figure*}

In the proton rest frame, for exclusive vector meson production in the dipole model representation \cite{Rezaeian:2013tka,Forshaw:2003ki,Watt:2007nr}, as shown in Fig.$\,\,$\ref{fig:DISandDVMP} (Right), even though the momentum transfer $\boldsymbol{\Delta}  \neq 0$ ($\rm t=-\boldsymbol{\Delta}^{2}$), the imaginary part of its amplitude can be similarly expressed as \cite{Kowalski:2006hc}

\begin{equation}
	\begin{aligned}
		&\mathcal{A}_{T, L}^{\gamma^{*} p \rightarrow V p}(x, Q^{2}, t)= \mathrm{i} \int \mathrm{d}^{2} \boldsymbol{r} \int_{0}^{1} \frac{\mathrm{d} z}{4 \pi} \int \mathrm{d}^{2}\boldsymbol{b}\\
		& \left(\Psi_{V}^{*} \Psi\right)_{T, L}\left(r,z,Q\right) \mathrm{e}^{-\mathrm{i}[\textbf{b}-(1-z)\textbf{r}]\boldsymbol{\Delta}}\cdot \frac{ \mathrm{~d} \sigma_{q \bar{q}}}{\mathrm{~d}^{2} \boldsymbol{b}}\left(x,r,b\right)\\
			&=\mathrm{i} \int_{0}^{\infty} \mathrm{d} r(2 \pi r) \int_{0}^{1} \frac{\mathrm{d} z}{4 \pi} \int_{0}^{\infty} \mathrm{d} b(2 \pi b) \\
			&\times\left(\Psi_{V}^{*} \Psi\right)_{T, L} J_{0}(b \Delta) J_{0}([1-z] r \Delta) \frac{\mathrm{d} \sigma_{q \bar{q}}}{\mathrm{~d}^{2} \boldsymbol{b}},
	\end{aligned}
\label{equation:VMP amplitude}
\end{equation}
where $\left(\Psi_{V}^{*} \Psi\right)_{T, L}$ is the amplitude of the conversion of virtual photon into vector meson. $\frac{ \mathrm{~d} \sigma_{q \bar{q}}}{\mathrm{~d}^{2} \boldsymbol{b}}$ is dipole scattering differential cross section where $\boldsymbol{b}$ is the impact parameter. Total dipole-target cross section $\sigma_{q \bar{q}}$ will be obtained by the BK equation. And $J_{0}$ is the  first kind Bessel function. To get the real part of the amplitude, the ratio of real to imaginary parts of the scattering amplitude will be used.

Thereby, differential cross section for exclusive vector meson
production is given by \cite{Nemchik:1994fp,Nemchik:1996cw,Forshaw:2003ki,Kowalski:2003hm,Kowalski:2006hc}
\begin{equation}
	\frac{\mathrm{d} \sigma_{T, L}^{\gamma^{*} p \rightarrow V p}}{\mathrm{~d} t}\left(x,Q^2,t \right)=\frac{R^2_g}{16 \pi}\left|\mathcal{A}_{T, L}^{\gamma^{*} p \rightarrow V p}\right|^{2}\left(1+\beta^{2}\right)
	\label{equation:Differential cross section}
\end{equation}
where $\beta$ denotes the ratio of real to imaginary parts of the scattering amplitude. $\beta$ is written as

\begin{equation}
	\beta=\tan (\pi \lambda / 2), \quad \text { with } \quad \lambda \equiv \frac{\partial \ln \left(\mathcal{A}_{T, L}^{\gamma^{*} p \rightarrow V p}\right)}{\partial \ln (1 / x)}.
	\label{equation:beta}
\end{equation}
And $R^2_g$ reflects the skewed effect, given by \cite{Shuvaev:1999ce} 

\begin{equation}
	R_{g}=\frac{2^{2 \lambda+3}}{\sqrt{\pi}} \frac{\Gamma(\lambda+5 / 2)}{\Gamma(\lambda+4)}.
	\label{equation:R_g}
\end{equation}

If the dependence of $t$ in the amplitude $\mathcal{A}_{T, L}^{\gamma^{*} p \rightarrow V p}$ is exponential \cite{Ahmady:2016ujw}, then Eq.$\,$ (\ref{equation:Differential cross section}) is rewritten as
\begin{equation}
	\frac{\mathrm{d} \sigma_{T, L}^{\gamma^{*} p \rightarrow V p}}{\mathrm{~d} t}\left(x,Q^2,t \right)=\frac{R^2_g}{16 \pi}\left|\mathcal{A}_{T, L}^{\gamma^{*} p \rightarrow V p}\right|_{t=0}^{2}\left(1+\beta^{2}\right)e^{-B_{D}|t|}.
	\label{equation:Differential cross sectiont=0}
\end{equation}

Then the total cross section is obtained as
\begin{equation}
	\sigma_{T, L}^{\gamma^{*} p \rightarrow V p}\left(x,Q^2\right)=\frac{R^2_g}{16\pi B_{D}}\left|\mathcal{A}_{T, L}^{\gamma^{*} p \rightarrow V p}\right|_{t=0}^{2}\left(1+\beta^{2}\right),
	\label{equation:tot cross section}
\end{equation}
\begin{equation}
	\sigma_{tot}^{\gamma^{*} p \rightarrow V p}\left(x,Q^2\right)=\sigma_{T}^{\gamma^{*} p \rightarrow V p}\left(x,Q^2\right)+\sigma_{L}^{\gamma^{*} p \rightarrow V p}\left(x,Q^2\right).
	\label{equation:Tot sigma }
\end{equation}
where $B_{D}=N\left(14.0\left(\frac{1 \mathrm{GeV}^{2}}{Q^{2}+M_{V}^{2}}\right)^{0.2}+1\right)$ with $N=0.55 \,\rm GeV^{-2}$ for $\rho^0$ \cite{Ahmady:2016ujw}. For $J/\Psi$, $B_D$ is written as \cite{Cepila:2019skb,ZEUS:2002wfj}
\begin{equation}
	B_D=
	\begin{cases}
		4.15+4\times0.116 \ln(\frac{\rm W}{90\,\rm GeV}) &Q^2 \leq 1 \rm GeV^2, \\
		4.72+4\times0.07 \ln(\frac{\rm W}{90\,\rm GeV}) &Q^2 > 1 \rm GeV^2,
	\end{cases}
\end{equation}
where $\rm W$ is center of mass energy of $\gamma^*p$, which is related to the $x$ and $Q^2$ as 

\begin{equation}
	x\,=\,x_{Bj}(1+M^2_V/Q^2)\,=\,\frac{Q^2+M^2_V}{W^2+Q^2},
\end{equation} 
where $M_V$ is the mass of the vector meson, and $x_{Bj}$ is bjorken scale. For $J/\Psi$, $M_V \,=\,3.097\, \rm GeV$, and for $\rho^0$, $M_V \,=\, 0.776 \,\rm GeV$.

For overlaps between photon and vector meson wave functions \cite{Kowalski:2006hc},
its transverse and longitudinal polarization parts are

\begin{equation}
	\begin{aligned}
		\left(\Psi_{V}^{*} \Psi\right)_{T}=& \hat{e}_{f} e \frac{N_{c}}{\pi z(1-z)}\left\{m_{f}^{2} K_{0}(\epsilon r) \phi_{T}(r, z)\right.\\
		&\left.-\left[z^{2}+(1-z)^{2}\right] \epsilon K_{1}(\epsilon r) \partial_{r} \phi_{T}(r, z)\right\}, \\
		\left(\Psi_{V}^{*} \Psi\right)_{L}=& \hat{e}_{f} e \frac{N_{c}}{\pi} 2 Q z(1-z) K_{0}(\epsilon r)\left[M_{V} \phi_{L}(r, z)\right.\\
		&\left.+\delta \frac{m_{f}^{2}-\nabla_{r}^{2}}{M_{V} z(1-z)} \phi_{L}(r, z)\right],
	\end{aligned}
\label{equation: wave function}
\end{equation}
where  $ e=\sqrt{4\pi \alpha_{em}}$, $N_c=3$, $\nabla_{r}^{2} \equiv(1 / r) \partial_{r}+\partial_{r}^{2}$, $\epsilon=\sqrt{z(1-z)Q^2+m^2_f}$, and the effective charge $\hat{e}_{f}$ is $2/3$ or $1/\sqrt{2}$ for $J/\psi$ or $\rho^0$. $m_f$ is the quark mass. $K_{0}$ and $K_{1}$ are the second kind Bessel function. For scalar wave functions, $\phi_{L}(r, z)$ and $\phi_{T}(r, z)$, there are two models, ``Gaus-LC" \cite{Kowalski:2003hm} and ``boosted Gaussian" \cite{Forshaw:2003ki}.  It should be noted that we follow the works of other people to use $\delta=0$ for the "Gaus-LC" model and $\delta=1$ for the "boosted Gaussian" model as mentioned by H.$\,$Kowalski et al \cite{Kowalski:2006hc}.

\begin{figure*}[htbp]
	\centering
	\includegraphics[scale=0.87]{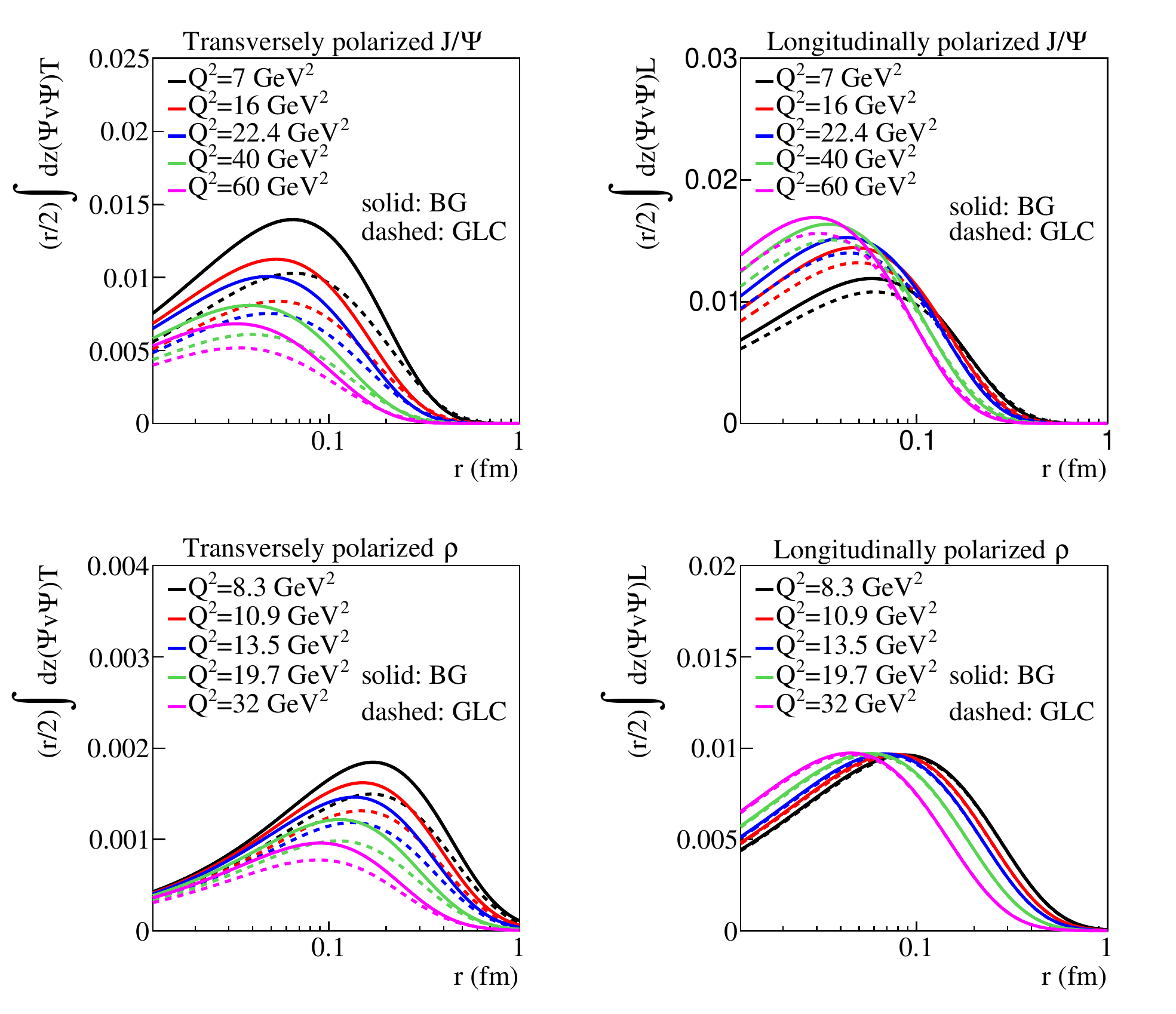}
	\caption{ Transversely and longitudinally overlaps between vector meson and the photon functions as a function of dipole size $\rm r$, which are integrated over $\rm z$ is presented. Solid line is for ``Boosted Gaussian'' (BG) model, and dashed line is for ``Gaus-LC'' (GLC) model.}
		\label{fig:wave_function}
\end{figure*}

For the ``Gaus-LC" model, the scalar wave functions, $\phi_{L}(r, z)$ and $\phi_{T}(r, z)$ are written as

\begin{equation}
	\begin{gathered}
		\phi_{T}(r, z)=N_{T}[z(1-z)]^{2} \exp \left(-r^{2} / 2 R_{T}^{2}\right), \\
		\phi_{L}(r, z)=N_{L} z(1-z) \exp \left(-r^{2} / 2 R_{L}^{2}\right).
	\end{gathered}
\label{equation:Gaus-LC}
\end{equation}
For the "boosted Gaussian" model, $\phi_{L}(r, z)$ and $\phi_{T}(r, z)$ are given by 
\begin{equation}
	\begin{aligned}
		\phi_{T}(r, z)=& \mathcal{N}_{T} z(1-z) \exp \left(-\frac{m_{f}^{2} \mathcal{R}^{2}}{8 z(1-z)}\right.\\
		&\left.-\frac{2 z(1-z) r^{2}}{\mathcal{R}^{2}}+\frac{m_{f}^{2} \mathcal{R}^{2}}{2}\right),\\
		\phi_{L}(r, z)=& \mathcal{N}_{L} z(1-z) \exp \left(-\frac{m_{f}^{2} \mathcal{R}^{2}}{8 z(1-z)}\right.\\
		&\left.-\frac{2 z(1-z) r^{2}}{\mathcal{R}^{2}}+\frac{m_{f}^{2} \mathcal{R}^{2}}{2}\right).
	\end{aligned}
\label{equation:boosted Gaussian}
\end{equation}
The parameters in the Eq.$\,\,$(\ref{equation:Gaus-LC}) and Eq.$\,\,$(\ref{equation:boosted Gaussian}) can be found in \cite{Kowalski:2006hc}. Transversely and longitudinally overlaps between vector meson functions and the photon function integrated over $\rm z$ as a function of dipole size $\rm r$ are presented in Fig. \ref{fig:wave_function}.

In the following calculation, both models ``Gaus-LC'' and ``boosted Gaussian'' will be used. We will compare the effects of two models on the calculation results.

\begin{figure*}[htbp]
	\begin{minipage}[htpb]{0.5\linewidth}
		\centering
		\includegraphics[scale=0.45]{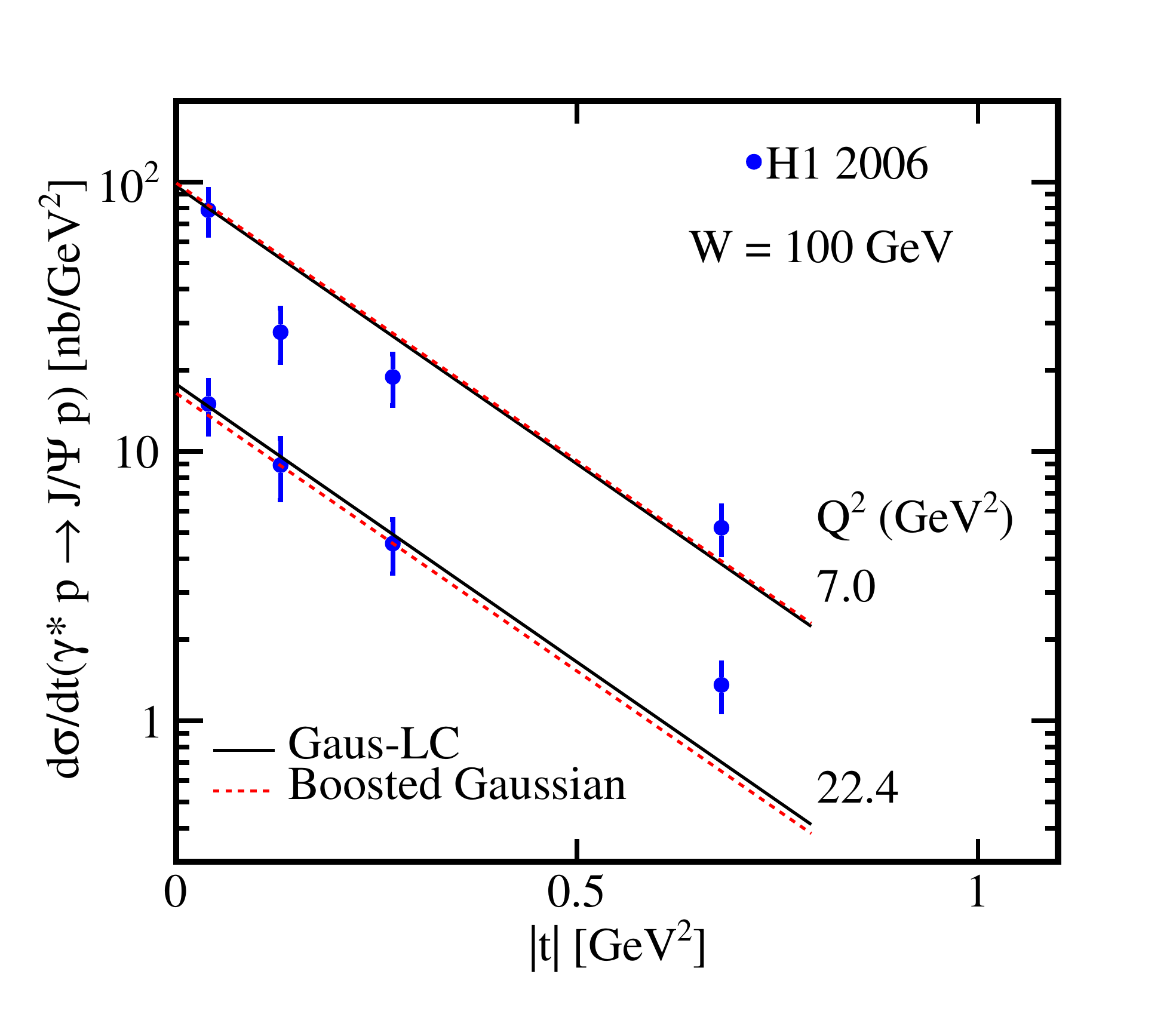}
	\end{minipage}%
	\begin{minipage}[htbp]{0.5\linewidth}
		\centering
		\includegraphics[scale=0.45]{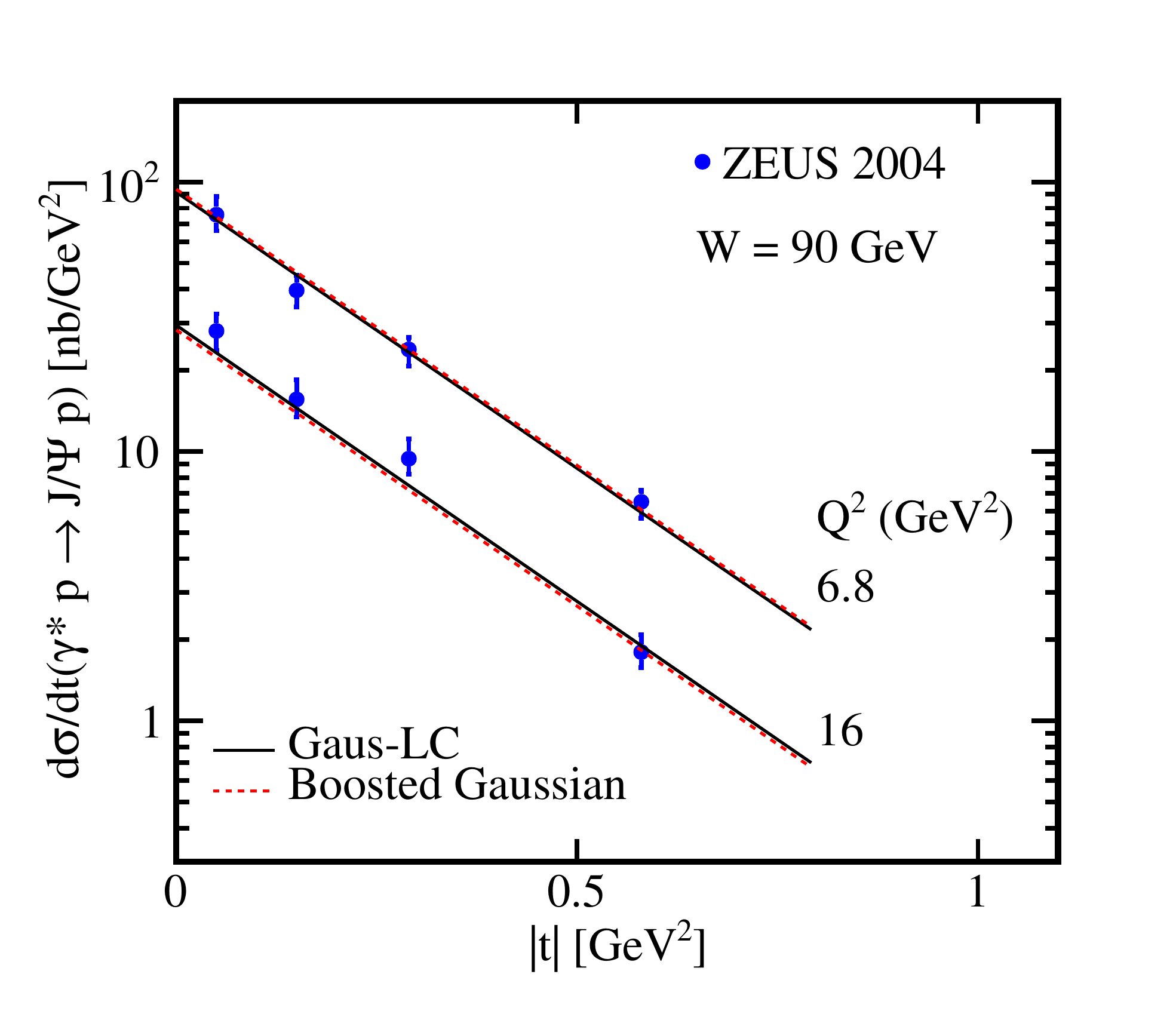}
	\end{minipage}
\begin{minipage}[htbp]{0.5\linewidth}
	\centering
	\includegraphics[scale=0.45]{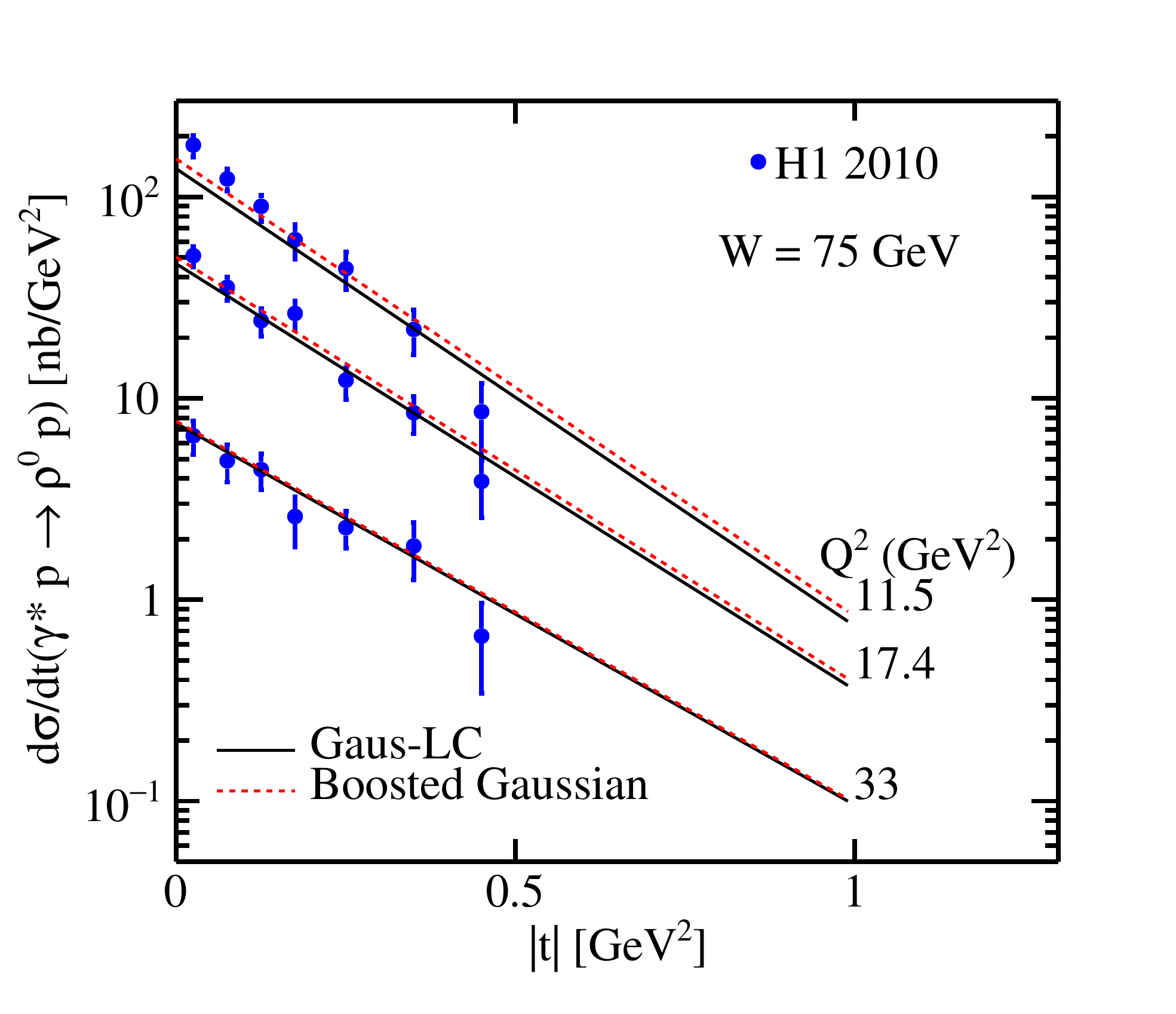}
\end{minipage}%
\begin{minipage}[htbp]{0.5\linewidth}
	\centering
	\includegraphics[scale=0.45]{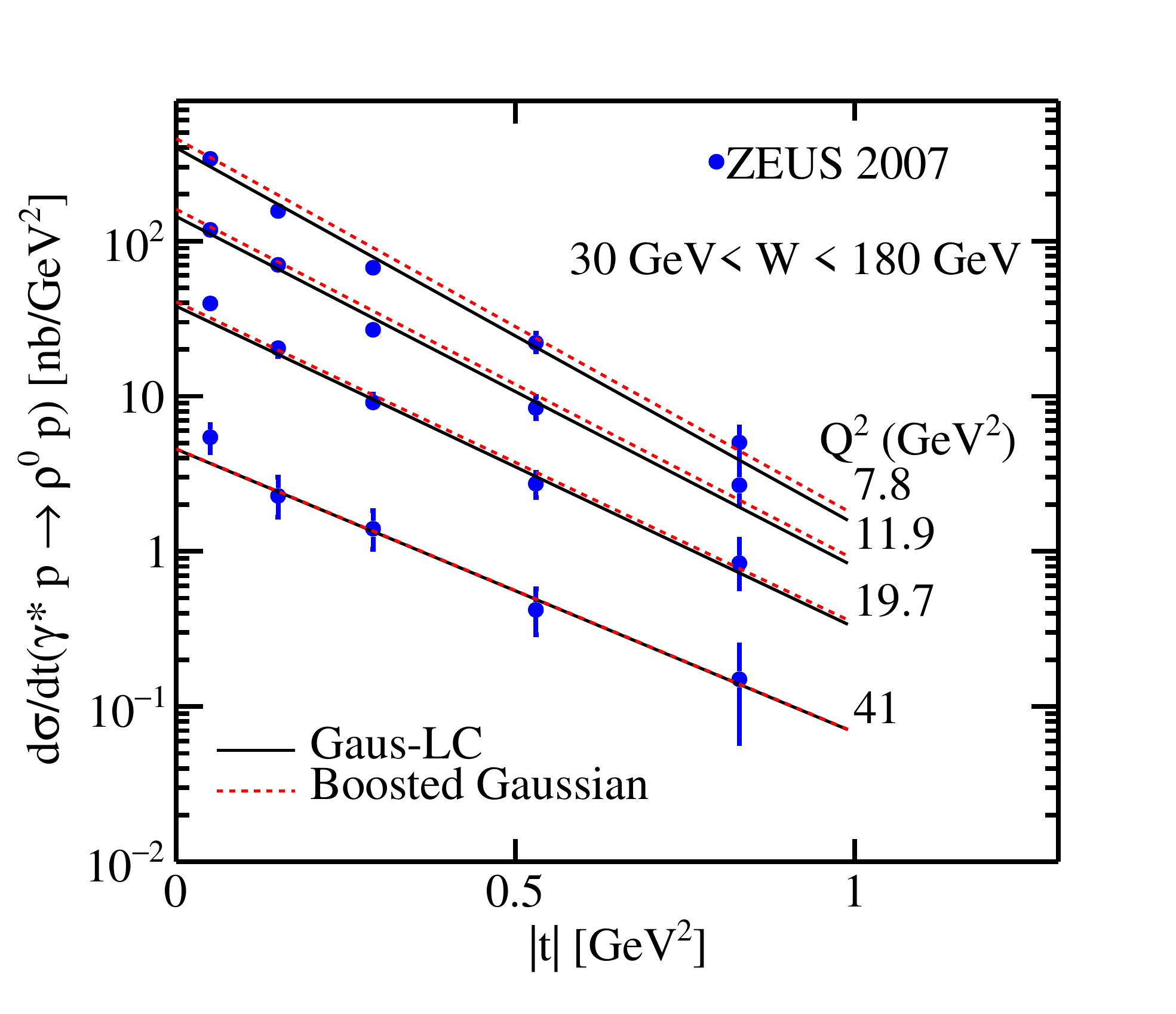}
\end{minipage}%
	\caption{ (above) Calculations of differential cross section of $\gamma^{*}p \rightarrow J/\Psi p$ as a function of $|\rm t|$ compared with experimental data from ZEUS 2004 \cite{ZEUS:2004yeh} and H1 2005 \cite{H1:2005dtp} using two different meson wave functions. (Bottom) Prediction of differential cross section of $\gamma^{*}p \rightarrow \rho^0 p$ as a function of $|\rm t|$ compared with experimental data from ZEUS 2007 and H1 2010  \cite{ZEUS:2007iet,H1:2009cml}.}	
	\label{fig:dsdt_J_Psi_rho_w}
\end{figure*}

\begin{figure*}[htbp]
	\begin{minipage}[t]{0.5\linewidth}
		\centering
		\includegraphics[scale=0.4]{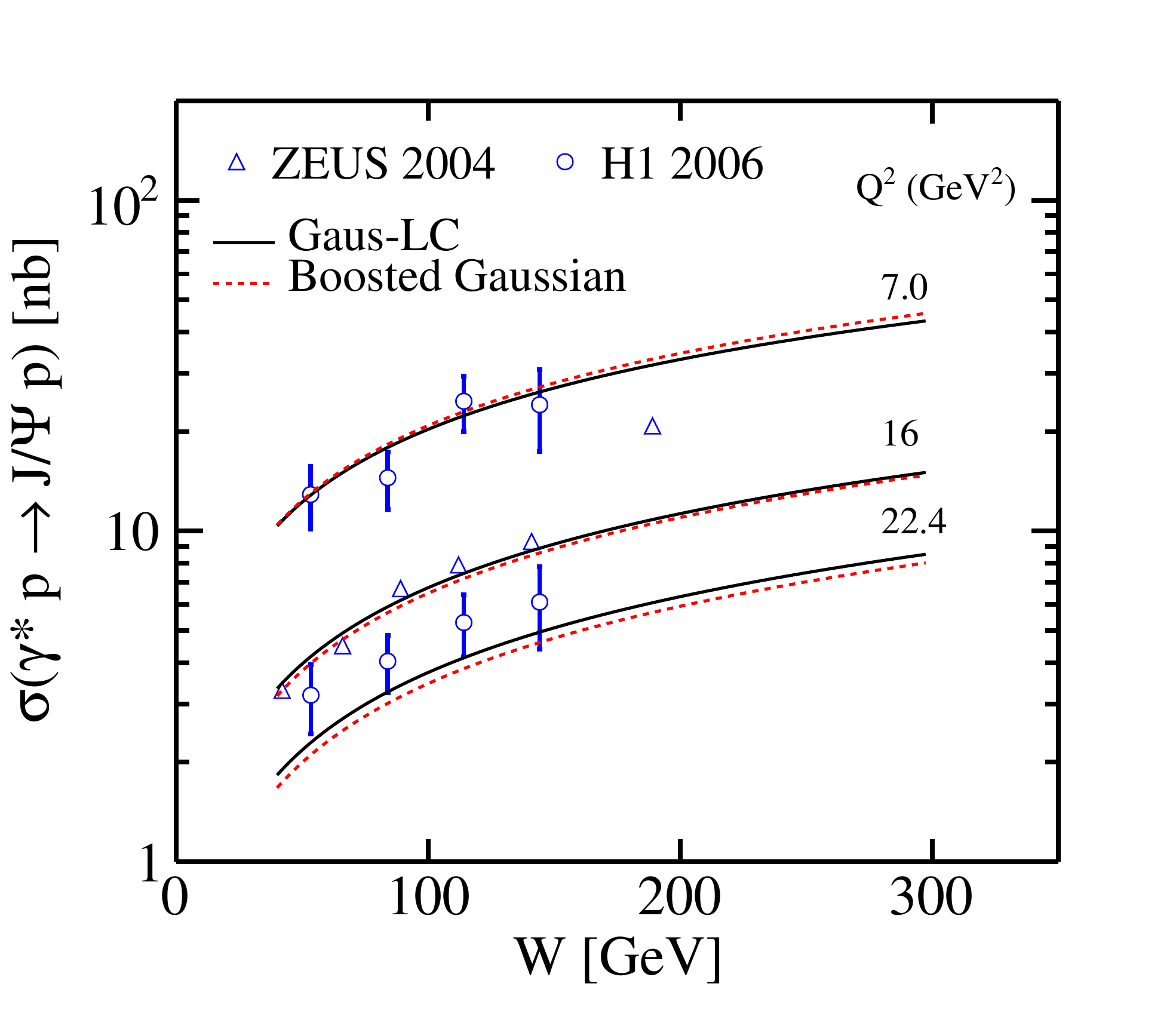}
	\end{minipage}%
	\begin{minipage}[t]{0.5\linewidth}
		\centering
		\includegraphics[scale=0.4]{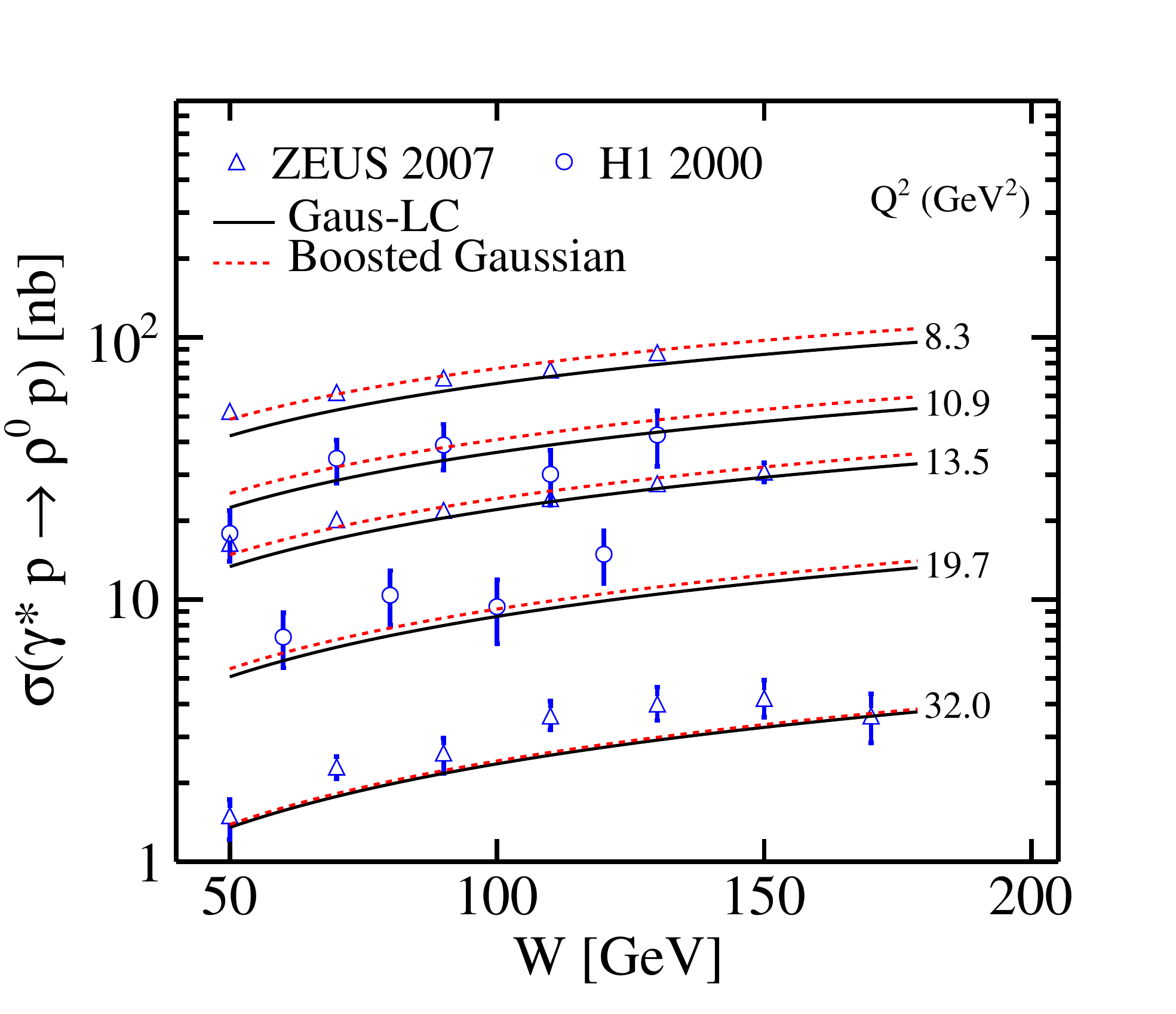}
	\end{minipage}
	\caption{ (Left) Prediction of total cross sections of $\gamma^{*}p \rightarrow J/\Psi p$ as a function of $\rm W$ compared with experimental data from ZEUS 2004 \cite{ZEUS:2004yeh} and H1 2006 \cite{H1:2005dtp} using two different meson wave functions. (Right) Prediction of total cross sections of $\gamma^{*}p \rightarrow \rho^0 p$ as a function of $\rm W$ compared with experimental data from H1 2000 \cite{H1:1999pji} and ZEUS 2007 \cite{ZEUS:2007iet}.}	
	\label{fig:J_Psi_rho_w}
\end{figure*}

\begin{figure*}[htbp]
	\begin{minipage}[t]{0.5\linewidth}
		\centering
		\includegraphics[scale=0.4]{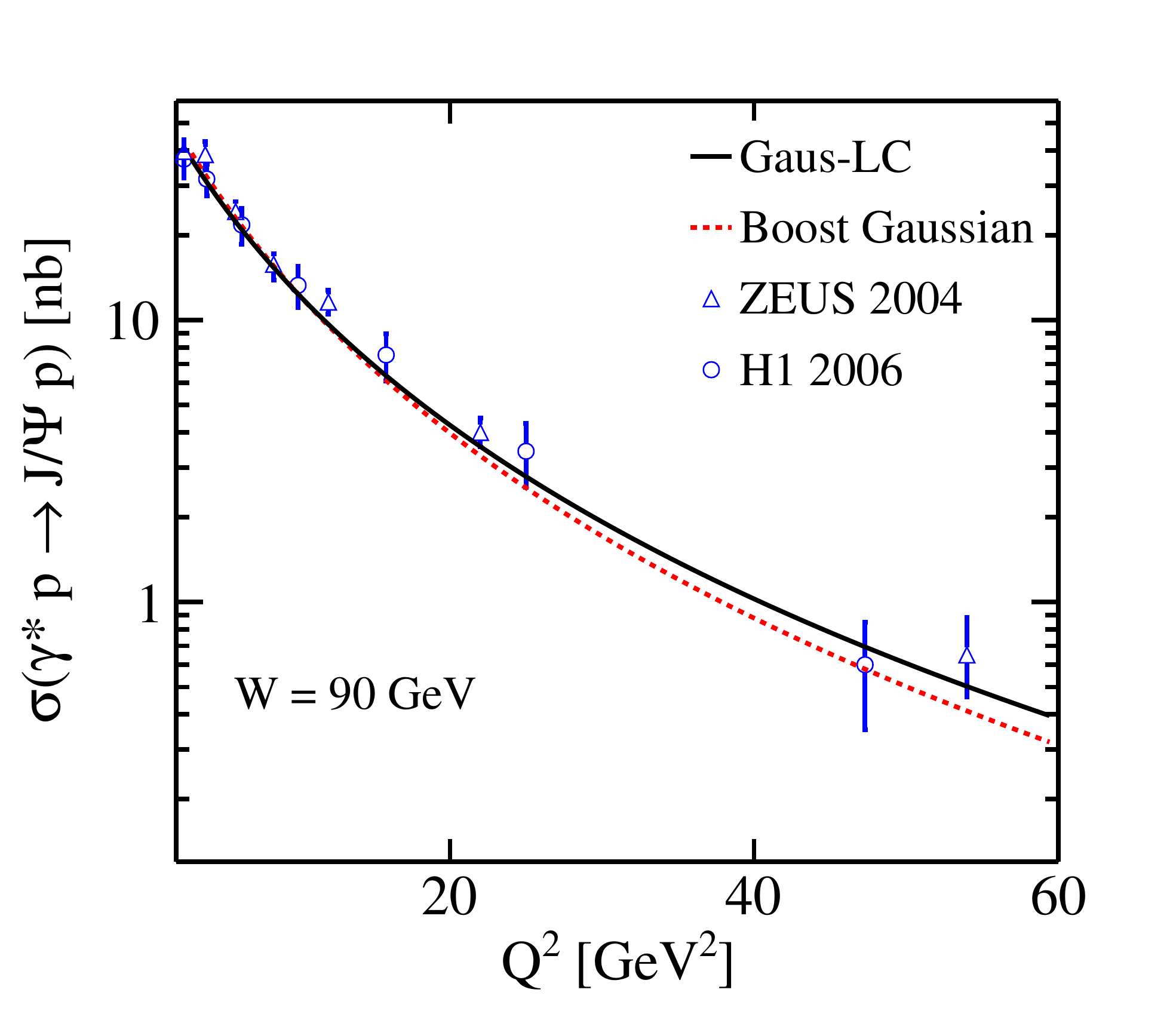}
	\end{minipage}%
	\begin{minipage}[t]{0.5\linewidth}
		\centering
		\includegraphics[scale=0.4]{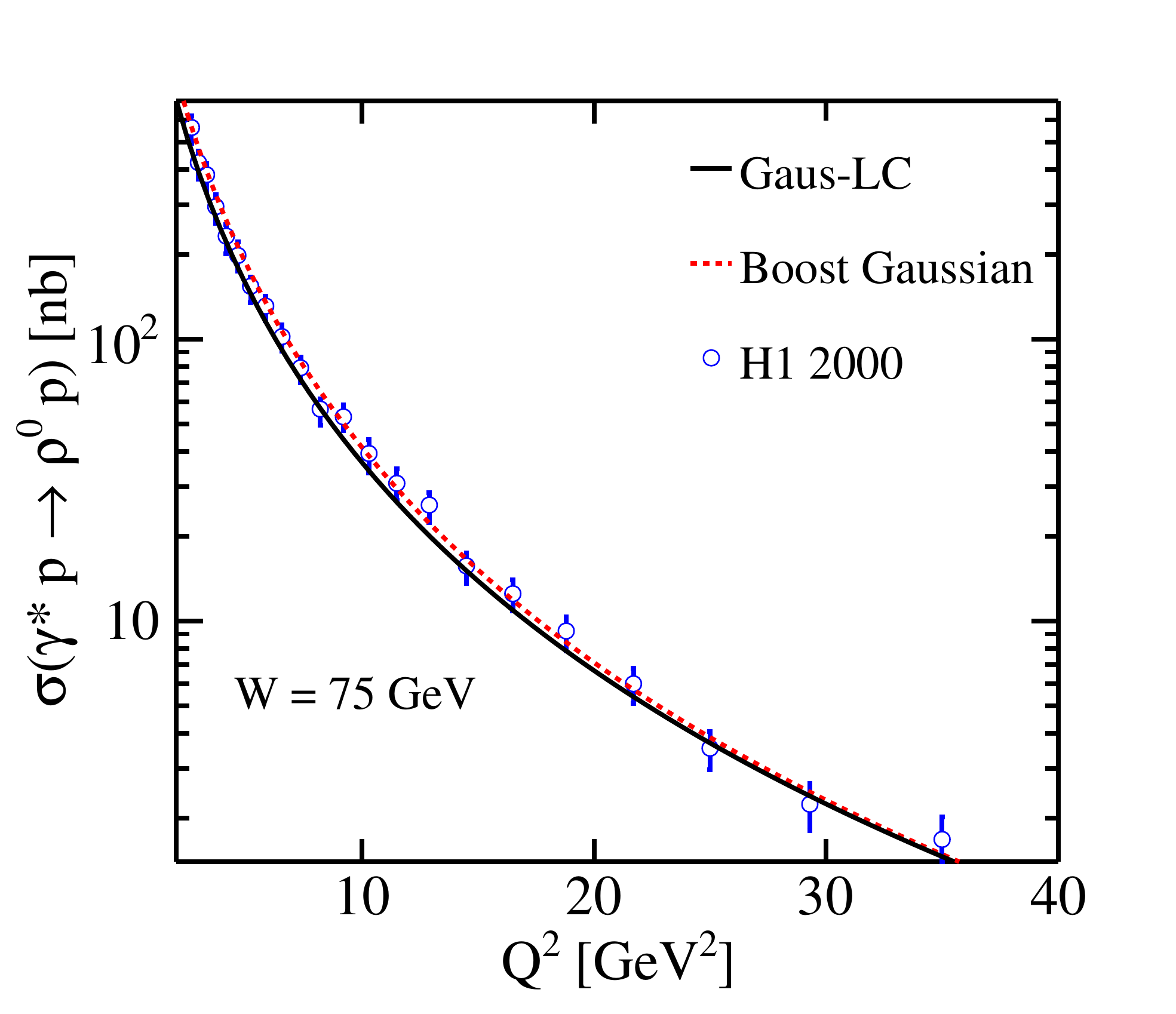}
	\end{minipage}
	\caption{(Left) Prediction for cross section of $\gamma^{*}p \rightarrow J/\Psi p$ as a function of $Q^2$ with two different meson wave functions compared with experimental data from ZEUS 2004 \cite{ZEUS:2004yeh} and H1 2006 \cite{H1:2005dtp}. (Right) Total cross sections of $\gamma^{*}p \rightarrow \rho^0 p$ as a function of $Q^2$ compared with experimental data from H1 2000 \cite{H1:1999pji} using two different meson wave functions.} 
	\label{fig:mesons_Q2}
\end{figure*}

\begin{figure*}[htbp]
	\begin{minipage}[t]{0.5\linewidth}
		\centering
		\includegraphics[scale=0.4]{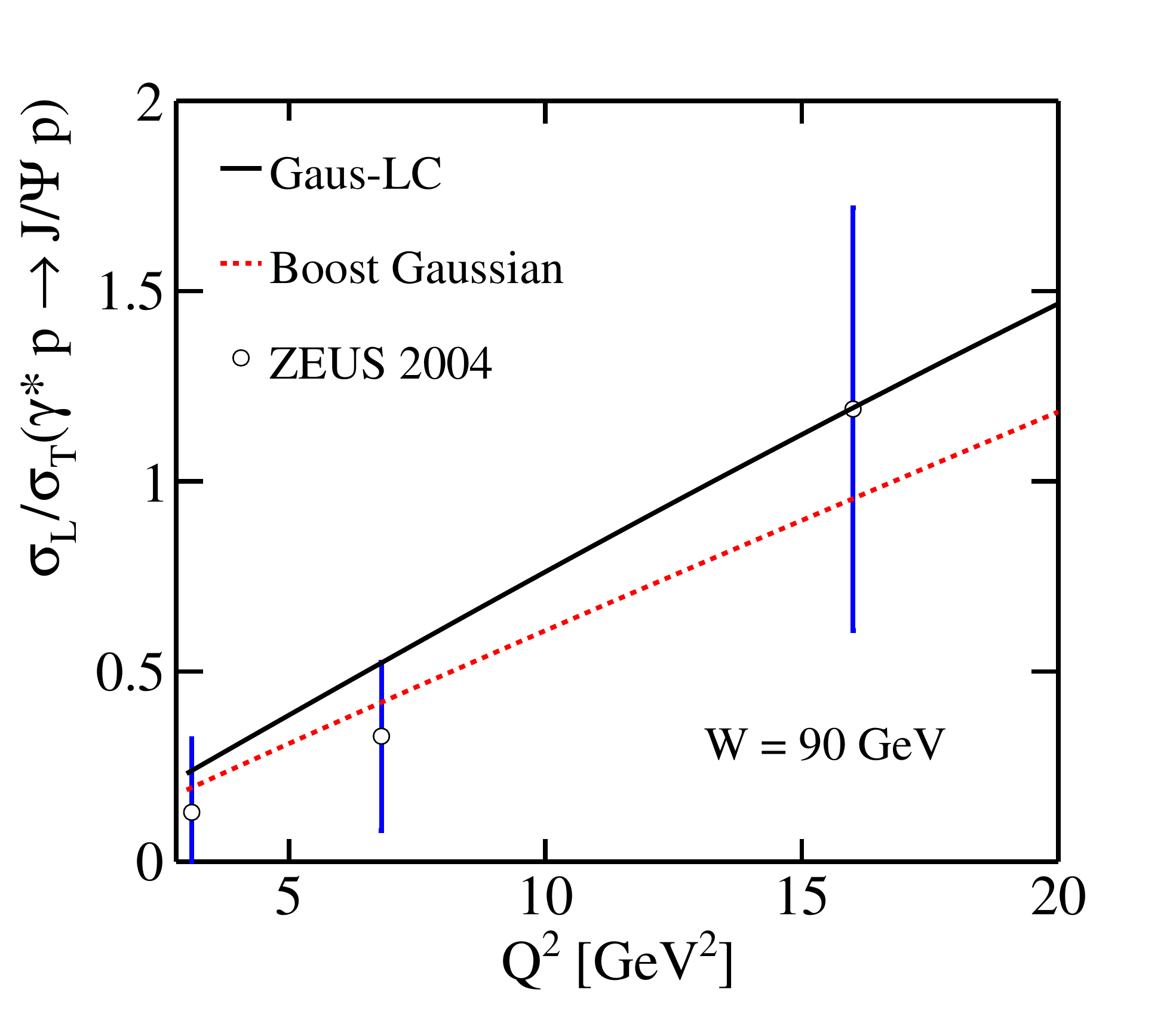}
	\end{minipage}%
	\begin{minipage}[t]{0.5\linewidth}
		\centering
		\includegraphics[scale=0.4]{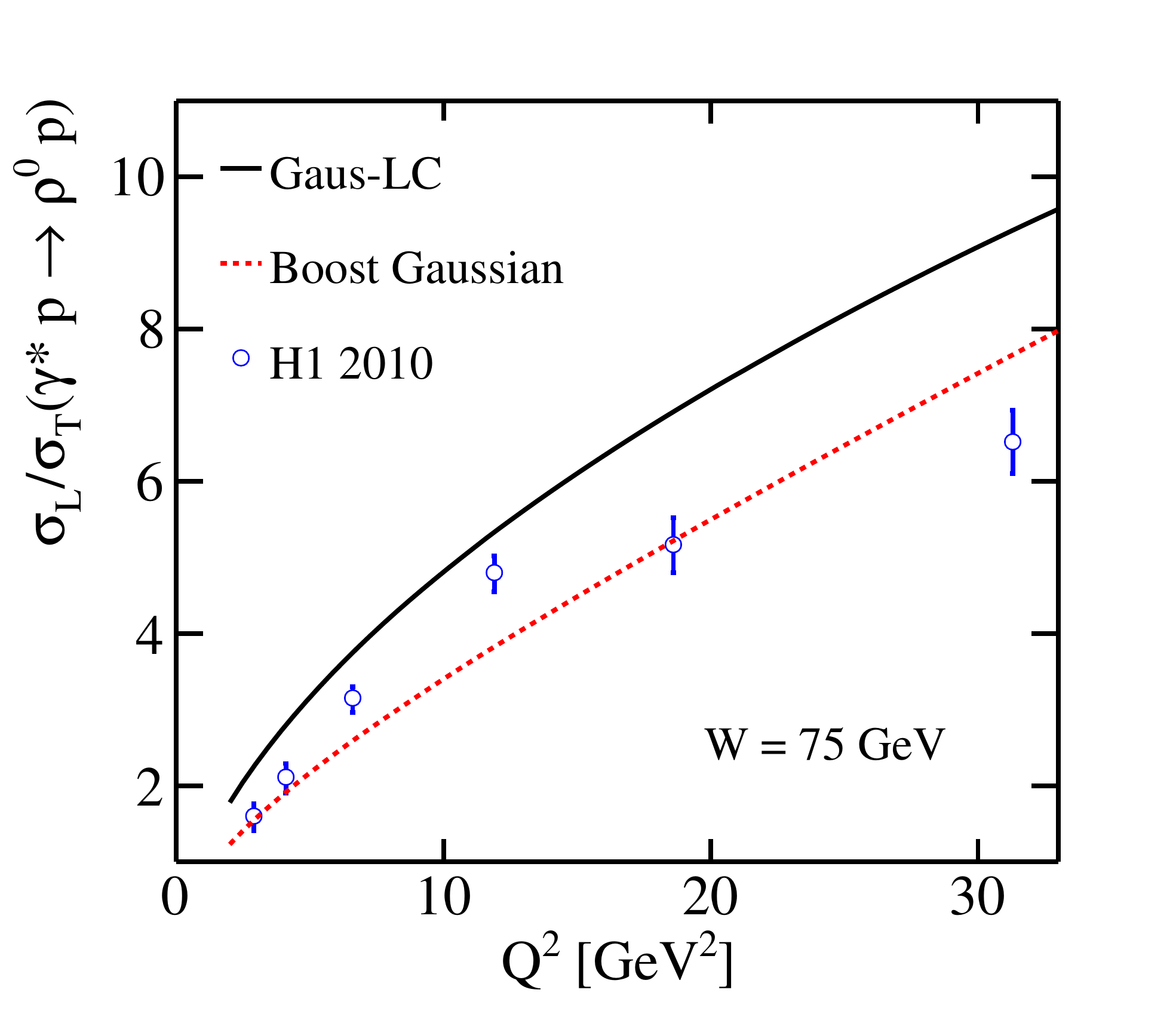}
	\end{minipage}
	\caption{(Left) For $\gamma^{*}p \rightarrow J/\Psi p$, the ratio of $\sigma_{L}$ and $\sigma_{T}$ as a function of $Q^2$ with two different meson wave functions compared with experimental data from ZEUS 2004 \cite{ZEUS:2004yeh} when $\rm W\,=\, 90\, \rm GeV$. (Right) For  $\gamma^{*}p \rightarrow \rho^0 p$, the ratio of $\sigma_{L}$ and $\sigma_{T}$ as a function of $Q^2$ compared with experimental data from H1 2010 \cite{H1:2009cml} using two different meson wave functions, when $\rm W\,=\, 75\, \rm GeV$.} 
	\label{fig:ratio}
\end{figure*}

\section{analytical solution of BK equation}
\label{sec:solution of BK equation}
In addition to wave functions, dipole-target scattering amplitude also determines exclusive vector meson production process.
In $\left|\mathcal{A}_{T, L}^{\gamma^{*} p \rightarrow V p}\right|_{t=0}$ of Eq.$\,\,$(\ref{equation:tot cross section}), dipole scattering differential cross section $ \mathrm{d} \sigma_{q \bar{q}}/\mathrm{d}^{2} \boldsymbol{b}$ is integrated and written as
\begin{equation}
	\int \frac{ \mathrm{~d} \sigma_{q \bar{q}}}{\mathrm{~d}^{2} \boldsymbol{b}}(x,r,\boldsymbol{b}) \mathrm{d}^{2} \boldsymbol{b}=\sigma_{q \bar{q}}(x,r)=2\pi R_p^{2}N(x,r),
\end{equation}
where $R_p$ is proton radius which will be obtained by fitting with structure function $F_2$ data of proton from DIS process, and the scattering amplitude $N(x,r)$ comes from BK equation. 

In the momentum space, with proper approximation, BK equation is represented as a  nonlinear evolution equation \cite{Munier:2003vc}
\begin{equation}
	\begin{aligned}
		\frac{\partial \mathcal{N}(\rm k, \rm Y)}{\partial \rm Y}=& \frac{\alpha_{\mathrm{s}} N_{\mathrm{c}}}{\pi} \chi\left(-\frac{\partial}{\partial \ln \rm k^{2}}\right) \mathcal{N}(\rm k, \rm Y) \\
		&-\frac{\alpha_{\mathrm{s}} N_{\mathrm{c}}}{\pi} \mathcal{N}^{2}(\rm k, \rm Y),
	\end{aligned}
\label{equation:BK equation}
\end{equation}
where $\chi(\lambda)=\psi(1)-\frac{1}{2} \psi\left(1-\frac{\lambda}{2}\right)-\frac{1}{2} \psi\left(\frac{\lambda}{2}\right)$ is BFKL kernel with $\psi(\lambda)=\Gamma^{\prime}(\lambda) / \Gamma(\lambda)$. In addition, $\rm Y=ln$$\frac{1}{x}$.

When fixing the running strong coupling constant and expanding the BFKL kernel $\chi(\lambda)$ \cite{Marquet:2005ic}, we obtain
\begin{equation}
	A_{0} \mathcal{N}-\mathcal{N}^{2}-\frac{\partial \mathcal{N}}{\partial \rm Y}-A_{1} \frac{\partial \mathcal{N}}{\partial \rm L}+\sum_{p=2}^{P}(-1)^{p} A_{p} \frac{\partial^{p} \mathcal{N}}{\partial^{p} \rm L}=0.
	\label{equation:expand BK equation}
\end{equation}
 Here, we need to clarify that the unit of $\rm Y$ is $\bar{\alpha}_s=\frac{\alpha_sN_c}{\pi}$, and $\rm L=\rm ln\,k^2/k^2_0$. For the calculation, we have fixed $\bar{\alpha}_s=0.191$ and $ \rm k^2_0=\Lambda^{2}_{\rm QCD}=0.04 \, \rm GeV^2$. With $P=2$, the simplified BK equation is given by
 \begin{equation}
 	A_{0} \mathcal{N}-\mathcal{N}^{2}-\frac{\partial \mathcal{N}}{\partial \rm Y}-A_{1} \frac{\partial \mathcal{N}}{\partial \rm L}+A_{2} \frac{\partial^{2} \mathcal{N}}{\partial^{2}\rm  L}=0.
 	\label{equation:simplified BK equation}
 \end{equation}
 The analytical solutions of Eq.$\,\,$(\ref{equation:simplified BK equation}) has be given as \cite{Wang:2020stj}
\begin{equation}
	\mathcal{N}(\rm L, Y)=\frac{A_{0} e^{5 A_{0}\rm Y / 3}}{\left[e^{5 A_{0} \rm Y / 6}+e^{\left[-\theta+\sqrt{A_{0} / 6 A_{2}}\left(\rm L- A_{1} \rm Y\right)\right]}\right]^{2}}.
	\label{equation:solution}
\end{equation}

 In this work, the values of $\rm A_0$, $\rm A_1$, $\rm A_2$ and $\theta$ are given by fitting to proton structure function $F_2$ with the following relationship between $F_2$ and $\mathcal{N}(L, Y)$ \cite{deSantanaAmaral:2006fe}:
 \begin{footnotesize}
\begin{equation}
	\begin{aligned}
			F_{2}\left(x, Q^{2}\right)=\frac{Q^{2} R_{p}^{2} N_{c}}{4 \pi^{2}} \int_{0}^{\infty} \frac{d \rm{k}}{\rm{k}} \int_{0}^{1} d z
			\left|\tilde{\Psi}_\gamma\left(\rm{k^{2}}, z ; Q^{2}\right)\right|^{2}
			\mathcal{N}(\rm L, Y),
		\label{equation:F_2}
	\end{aligned}
\end{equation}
\end{footnotesize}
where $N_{c}$ is the number of colors. Here $x\,=\,x_{Bj}$. As shown in Fig.$\,$\ref{fig:DISandDVMP} (Left), the wave function $\left|\tilde{\Psi}_\gamma\left(\rm{k^{2}}, z ; Q^{2}\right)\right|^{2}$ expressed in momentum space, represents the probability of a virtual photon splitting into a quark-antiquark pair. It is given by \cite{deSantanaAmaral:2006fe} 

\begin{footnotesize}
	\begin{equation}
		\begin{aligned}
			\left|\tilde{\Psi}_\gamma\left({\rm k^{2}}, z ; Q^{2}\right)\right|^{2}&= \sum_{q}\left(\frac{4 \bar{Q}_{q}^{2}}{{\rm k^{2}}+4 \bar{Q}_{q}^{2}}\right)^{2} e_{q}^{2}\left.\bigg \{ \left[z^{2}+(1-z)^{2}\right]\right.\\
			&\times\left[\frac{4\left({\rm k^{2}}+\bar{Q}_{q}^{2}\right)}{\sqrt{{\rm k^{2}}\left({\rm k^{2}}+4 \bar{Q}_{q}^{2}\right)}} \operatorname{arcsinh}\left(\frac{{\rm k}}{2 \bar{Q}_{q}}\right)\right.\\
			&\left.+\frac{{\rm k^{2}}-2 \bar{Q}_{q}^{2}}{2 \bar{Q}_{q}^{2}}\right]+\frac{4 Q^{2} z^{2}(1-z)^{2}+m_{q}^{2}}{\bar{Q}_{q}^{2}} \\
			& \times\left[\frac{{\rm k^{2}}+\bar{Q}_{q}^{2}}{\bar{Q}_{q}^{2}}-\frac{4 \bar{Q}_{q}^{4}+2 \bar{Q}_{q}^{2} {\rm k^{2}+k^{4}}}{\bar{Q}_{q}^{2} \sqrt{{\rm k^{2}}\left({\rm k^{2}}+4 \bar{Q}_{q}^{2}\right)}}\right.\\
			&\left.\left.\times \operatorname{arcsinh}\left(\frac{{\rm k}}{2 \bar{Q}_{q}}\right)\right]\right.\bigg \},
		\end{aligned}
		\label{equation:wave_function_DIS}
	\end{equation}  
\end{footnotesize}
 
where $\bar{Q}_{q}^{2}=z(1-z) Q^{2}+m_{q}^{2}$, $m_{q}$ the mass of the quark of flavor $q$.  

 In \cite{deSantanaAmaral:2006fe}, their fitting range for $Q^2$ is $0.045 \, \leq Q^{2}  \leq 150 \, \mathrm{GeV}^{2}$, because the corrections from DGLAP equation should be considered in too high $Q^2$ range.  So the kinematic fitting range we choose is
 \begin{equation}
 	\begin{gathered}
 		x \, \leq \, 0.01, \\
 		1  \,\mathrm{GeV}^{2}\, < \, Q^{2} \,\leq \, 45 \, \mathrm{GeV}^{2}.
 	\end{gathered}
 \label{eq:fit_range}
 \end{equation} 
 
 The fitting results are presented in Fig.$\,$$\ref{fig:F19},\,\ref{fig:F20}$. the values of the parameters obtained by fitting to 85 $F_2$ data points from H1 and ZEUS \cite{H1:2009pze,H1:2013ktq} with $m_{u,d,s}\,=\,0.14\,\rm GeV$ and $m_c\,=\,1.4\,\rm GeV$ are listed in table.$\,$\ref{tab:fit_par}. The $\chi^2$ is also provided in table.$\,$\ref{tab:fit_par} .
\begin{table*}
	\centering
	\caption{Parameters from the fit to the $F_2$ data \cite{H1:2009pze,H1:2013ktq} and $\chi^2$ per data points (nop means ``number of points''). }	
	\label{tab:parameter of wave function}       %
	\resizebox{\linewidth}{!}{
	\begin{tabular}{cccccccccccc}
		
		\hline\noalign{\smallskip}
		\centering
		$x$&$Q^2/\rm\,GeV^{2}$&$m_{u,d,s}/\,\rm GeV$&$m_{c}/\,\rm GeV$&$\alpha_s$&$\rm k_0/\,\rm GeV$&$\rm A_0$ & $\rm A_1$ & $\rm A_2$ &$\theta$ &$R_p/\,\rm GeV^{-1}$ &$\chi^2/{\rm nop}$  \\
		
		\noalign{\smallskip}\hline\noalign{\smallskip}
		\centering
		$\leq\,0.01$&$(1,45]$&0.14&1.4&0.2&0.2&$0.696\,\pm\,0.0187$ & $0.661\,\pm\,0.0295$ & $0.112\,\pm\,8.0\times10^{-3}$ & $-0.463\,\pm\,0.0493$ &$5.484\,\pm\,0.106$ &0.979 \\
		
		
		\noalign{\smallskip}\hline	
	\end{tabular}
}
\label{tab:fit_par}
\end{table*}

Then in the following calculation, we need the dipole scattering amplitude $\mathcal{N}(x,\textbf{r})$ in the coordinate space. $\mathcal{N}(x,\textbf{r})$ is related to $\mathcal{N}(x,\textbf{k})$ by the Fourier transformation,
\begin{equation}
	\mathcal{N}(x,\textbf{k}) = \frac{1}{2\pi}\int\frac{d^2\textbf{r}}{r^2}e^{i\textbf{k}\cdot\textbf{r}}\mathcal{N}(x,\textbf{r}).
\label{equation:Fourier transform}
\end{equation}
By inverse Fourier transformation, we can get 
\begin{equation}
	\begin{aligned}
	\mathcal{N}(x,\textbf{r}) &= \frac{r^2}{2\pi}\int{d^2\textbf{k}}e^{-i\textbf{k}\cdot\textbf{r}}\mathcal{N}(x,\textbf{k})\\
	& =r^2\int^{\infty}_{0}dkkJ_0(k\cdot r)\mathcal{N}(x,k).
	\end{aligned}
\label{equation: Fourier inversion}
\end{equation}

Then, we apply Eq.$\,\,$(\ref{equation: Fourier inversion}) to the total cross section calculations of $J/\Psi$ and $\rho^0$ productions.

\section{Numerical results}
\label{sec:Numerical results}
The cross sections of vector mesons are calculated with the dipole-amplitude and overlaps between the vector meson and 
the photon wave functions. In this work, the wave functions of vector mesons are adopted in two models. The solution of BK equation in previous section is employed for dipole-amplitude. We examine  whether the solution of BK equation is valid in the calculations.

 We present the differential cross sections, total cross sections and ratios of $\sigma_{L}$ and $\sigma_{T}$  of $\rho^0$ and $J/\Psi$ productions respectively using our analytic solution of BK equation. For the wave functions of the total cross section, we consider the influence of the ``Gaus-LC'' model and the ``boosted Gaussian'' model respectively. Finally, the total cross sections are calculated theoretically and compared with the experimental data.

First of all, the results of differential cross sections of  $J/\Psi$ and $\rho^0$ productions as a function of t are presented compared with experiment data \cite{H1:2005dtp,ZEUS:2004yeh,ZEUS:2007iet,H1:2009cml}, as shown in Fig.$\,\,$\ref{fig:dsdt_J_Psi_rho_w}. Then
we predict the $\rm W$-dependence total cross sections of two meson in two wave functions models with our solutions of BK equation.
As shown in Fig.$\,\,$\ref{fig:J_Psi_rho_w}, for exclusive $J/\Psi$ and $\rho^0$ productions, we compare our results with the experimental data \cite{ZEUS:2004yeh,H1:2005dtp,H1:1999pji,ZEUS:2007iet}, in the case of $7.0 \, \rm GeV^2 \leq Q^2\leq22.4 \,\rm GeV^2$ and $8.3 \,\rm GeV^2\leq Q^2 \leq 32.0 \,\rm GeV^2$. From the results, the calculations are reasonable within the range of uncertainty allowed. And we can find the results obtained by ``Gaus-LC'' model and ``boosted Gaussian'' model become closing to each other as $ Q^2$ increasing. For $J/\Psi$, total cross sections obtained by two models are even flip-over at higher $ Q^2$. These are mainly from the contribution of the wave functions. As shown in Fig.$\,$\ref{fig:wave_function}, for transversely polarized part of $\rho^0$ and $J/\Psi$, the difference between the two models is gradually reduced as $ Q^2$ increases. For longitudinally polarized part of $\rho^0$, the two models are basically the same. But for $J/\Psi$, There are some differences between the two models.  

Secondly, the total cross sections as a function of $Q^2$ are presented in Fig.$\,\,$\ref{fig:mesons_Q2} when $\rm W=90\,\rm GeV$ for $J/\Psi$ and $\rm W=75\,\rm GeV$ for $\rho^0$. It can be seen that the predictions agree with 
the experimental data well. Then, to summarize, the solution of BK equation is valid to perform the cross sections of vector meson in diffractive process.

Thirdly, Fig.$\,\,$\ref{fig:ratio} shows the ratios of the longitudinal to transverse cross section, $\sigma_{L}/\sigma_{T}$ as a function of $Q^2$ for fixed $\rm W$. For $J/\Psi$ production, $\sigma_{L}/\sigma_{T}\,\propto Q^2$. And for $\gamma^*p\rightarrow \rho^0 p$ process, $\sigma_{L}/\sigma_{T}$ based on two wave function models grows rapidly with $Q^2$. As $Q^2$ increases, the ratio calculated by us exceeds the experimental data. In \cite{Kowalski:2006hc}, a reasonable explanation is given for this situation. The ratios of the longitudinal to transverse cross section are sensitive to the wave functions.

 Briefly, when the running coupling constant $\alpha_s$ is fixed, the solution of BK equation with the parameters fitted by us in this work is efficient in the prediction of the meson production. As $Q^2$ increases , $\alpha_s$ vanishes asymptotically. Only at large $Q^2$, we can assume that its dependence on $Q^2$ diminishes. Accordingly,  the $F_2$ data of $Q^2\,<\,1\,\rm GeV^2$ is not considered. And too high $Q^2$ data is also not taken into account, because the DGLAP corrections could work in too high $Q^2$ region \cite{deSantanaAmaral:2006fe}.

\section{ Discussion and Summary}
\label{sec:summary}
In this work, the good values of parameters in the solution of BK equation are obtained by fitting the $F_2$ data firstly. Then we use the solution of BK equation and two wave function models (the ``Gaus-LC'' model and the ``boosted Gaussian'' model) to get the total cross sections of the exclusive $\rho^0$ and $J/\psi$ productions and the ratios of the longitudinal to transverse cross section. The results show that the analytical solution of BK equation with fixed $\alpha_s$ is suitable within a certain $Q^2$ region. And the difference of the results caused by the two wave function models decreases with the increase of $Q^2$ for $\rho^0$. But for $J/\Psi$, at higher $Q^2$, the difference still occurs. 

To be more precise, our solution of BK equation with the parameters fitted by us is valid within not too high $Q^2$ region. If this region is exceeded, it may be more appropriate to use the DGLAP equation or improved BK equation \cite{Albacete:2015xza,Albacete:2009fh}. 

 In a word, vector meson production is effective to measure the properties of proton. In the future, EIC \cite{AbdulKhalek:2021gbh} and EicC \cite{Anderle:2021wcy} experimental data will offer more tests on BK equation.

\begin{acknowledgments}
This work is supported by
the Strategic Priority Research Program of Chinese Academy of Sciences under the Grant NO. XDB34030301.

\end{acknowledgments}

\begin{figure*}[htbp]
	\includegraphics[scale=0.9]{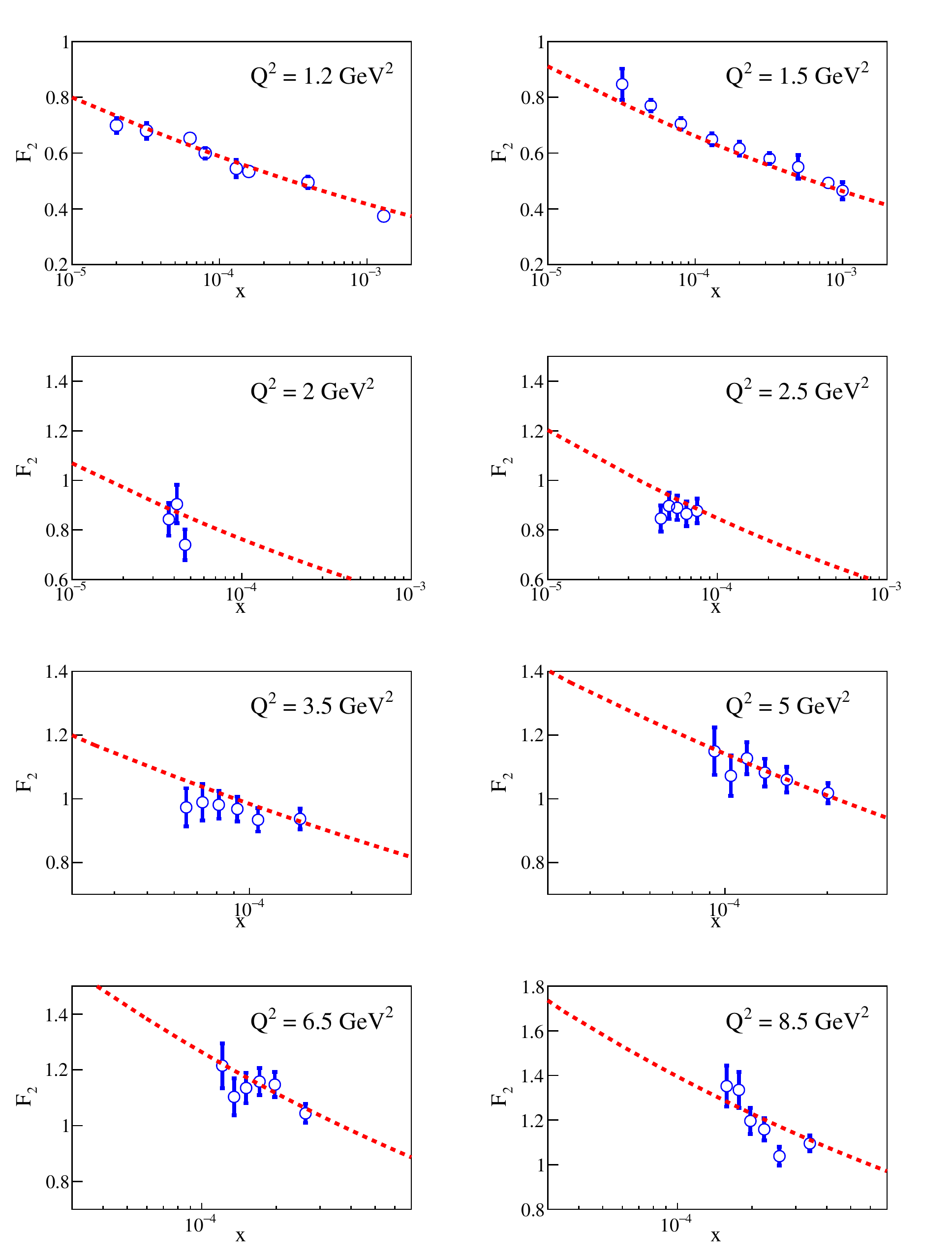}
	\caption{Result for fitting to proton structure function $ F_2$ as a function of $x$ at different $Q^2$. These data are from H1 and ZEUS \cite{H1:2009pze,H1:2013ktq}. The red dotted line is the fitting curve by eq.$\,$(\ref{equation:F_2}).}
	\label{fig:F19}
\end{figure*}

\begin{figure*}[htbp]
	\includegraphics[scale=0.9]{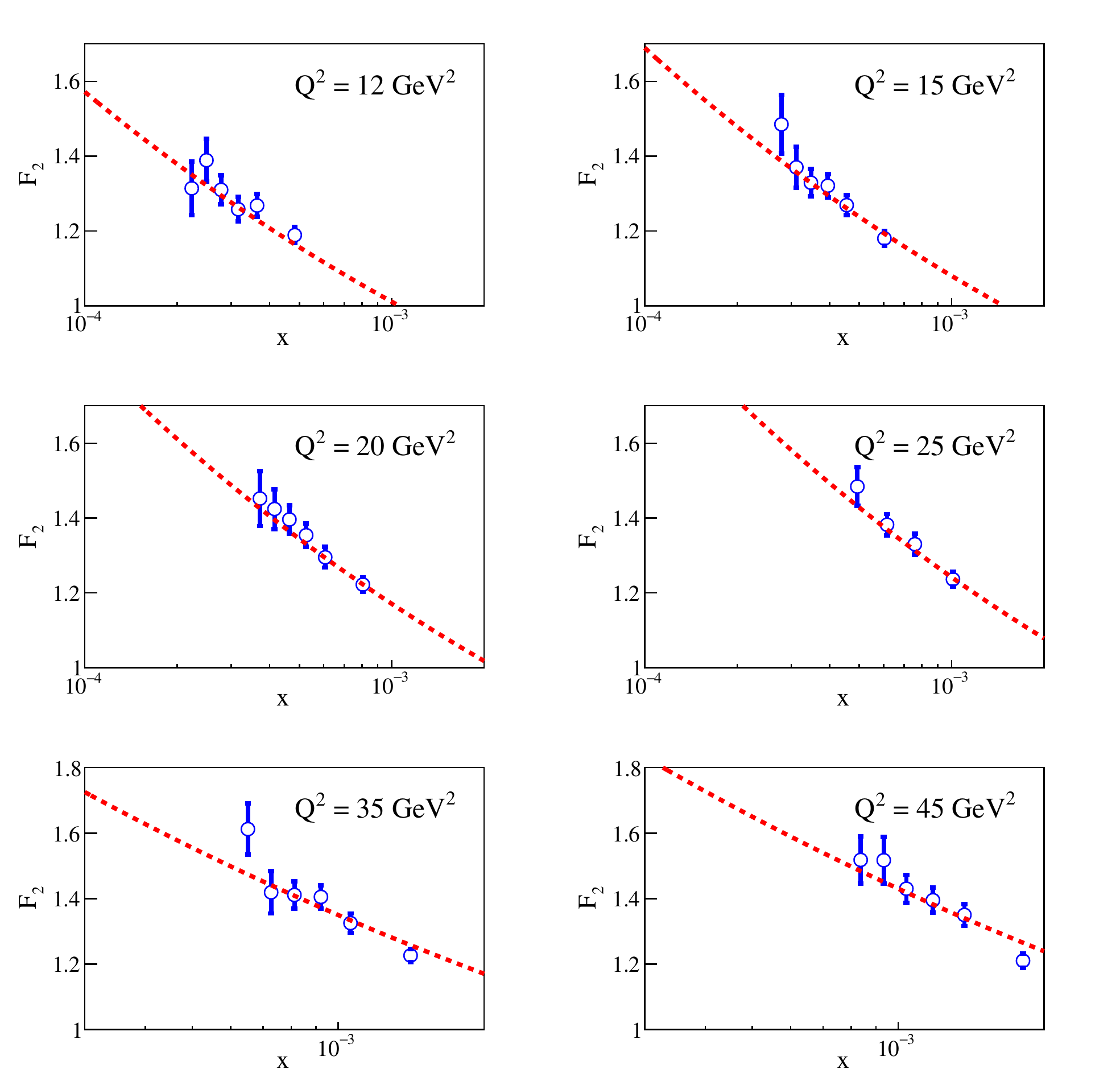}
	\caption{Continuation of Fig.$\,$\ref{fig:F19}.}
	\label{fig:F20}
\end{figure*}

\bibliographystyle{apsrev4-1}
\bibliography{refs_INSPIRE}

\end{document}